\documentclass[11pt]{article}      
\author{Kris Krogh \\
Neuroscience Research Institute\\ University of California, Santa
Barbara, CA 93106, USA\\ email: k\_krogh@lifesci.ucsb.edu}

\title
{Galactic Nuclei and Jets in Wave Gravity}

\date{June 24, 2006}

\begin{document}

\maketitle

\begin{abstract}

\normalsize  ``Wave gravity" refers to a quantum-mechanical gravity
theory \linebreak introduced in two previous papers~\cite{kk1,kk2}.
Although based on the optics of de Broglie waves instead of curved
space-time, it agrees with the standard tests of general relativity.
As in that theory, galactic nuclei are dark objects where gravity
prevents the escape of most radiation. In this case, collapse is
counteracted by rising internal pressure and black hole
singularities don't occur. Unlike black holes, these nuclei can have
internal magnetic fields, and high-energy plasma can escape along
magnetic field lines closely aligned with the gravitational field
direction. This allows a different model of jets from active
galactic nuclei, where jets can arise without direct fueling by
accretion disks.  It also offers a new basis for the tight
correlation observed~\cite{lf} between the masses of galactic nuclei
and their hosts.
\end{abstract}

\vfill\eject

\section{Introduction}

Previous papers~\cite{kk1,kk2} presented a new quantum-mechanical
theory of gravity, based on the optics of de Broglie waves rather
than curved space time.  Here we'll call it ``wave gravity."  This
paper derives its predictions for supermassive bodies, and compares
those to the observed properties of galactic nuclei.

As emphasized by Feynman~\cite{fls}, quantum-electrodynamics can be
described simply in terms of electromagnetic potentials and their
effects on the phases of quantum-mechanical waves.   Gravitational
dynamics have a similar basis in this theory.  Gravitational
potentials resemble the scalar potential in electromagnetism, and
are coupled to matter as
\begin{equation}
\Phi \:=\, \frac{-\;G\,m_0}{\sqrt{\,x^2 + (y^2 +z^2)(1-v^2/\,c^2)}}
\label{eq:1}
\end{equation}
Here $\Phi$ is the gravitational potential, $G$ the gravitational
constant, $m_0$ the inertial rest mass of a small mass element, and
$v$ is its velocity in the $x$ dimension of $xyz$ coordinates. These
potentials are also governed by a wave equation analogous to that
for the electromagnetic scalar potential
\begin{equation}
\nabla^2 \Phi \,-\, \frac{1}{c^2}\frac{\partial^2\Phi}{\partial t^2}
\:=\: 4 \pi G \rho
\end{equation}
where $\rho$ is the density of the mass $m_0$ appearing in the
previous equation.

The effect of gravitational potentials in this theory is a slowing
of quantum mechanical waves.  Space and time are treated as in the
preferred-frame special relativity advocated by Lorentz,
Poincar\'{e}, and Bell~\cite{jsb}.  (Where general relativity
assumes an absolute speed of light, and variable space and time,
this assumes the opposite.) In a gravitational potential, the speed
of light varies as
\begin{equation}
c \:=\: c_0 e^{2 \Phi / c_0^2} \label{eq:3}
\end{equation}
where $c_0$ is the value in the absence of a gravitational
potential.

De Broglie discovered the velocity $V$ of matter waves associated
with any massive particle is given by
\begin{equation}
V \:=\: \frac{\,c^2}{\,v}
\end{equation}
where $v$ is the particle's velocity.  Here $V$ diminishes in a
gravitational potential by the same factor as $c$, as
\begin{equation}
V \:=\: V_0 e^{2 \Phi / c_0^2}
\end{equation}
This gives gravity's influence on the motion of matter.

For a full description of the theory, see~\cite{kk1}.  There it's
derived from general principles and shown to agree with the standard
experimental tests of general relativity. A second paper~\cite{kk2}
shows that, unlike general relativity, it predicts the observed
motions of the Pioneer 10 and 11 space probes. (Those papers also
describe the associated cosmology, which has no ``flatness problem"
or ``age problem," and doesn't require strange dark matter or
energy.)

\section{Escape of Radiation from Supermassive Bodies}

From Eqs.~(\ref{eq:1}) and (\ref{eq:3}), the speed of light near a
stationary spherical body can be expressed as
\begin{equation}
c \:=\: c_0 e^{-2 \mu /r} \label{eq:6}
\end{equation}
where $r$ is the radial distance in isotropic coordinates and $\mu$
is defined as
\begin{equation}
\mu \:=\: \frac{G\,m_0}{c_0^2}
\end{equation}
This gives a spherically symmetric refractive index
\begin{equation}
n \:=\: \frac{\,c_0}{\,c} \:=\: e^{2\mu /r} \label{eq:8}
\end{equation}

In polar coordinates, the path of a light ray in a spherically
symmetric refractive index gradient is given by
\begin{equation} \theta \:=\: \theta_0 +
k\int_{r_0}^r \frac{dr}{r\,\sqrt{r^2 n^2 - \,k^2 }}
\end{equation}
where $\theta_0$ and $r_0$ are the coordinates of an arbitrary
starting point~\cite{ewm}. The quantity $k$ is a constant given by
\begin{equation}
k \:=\: \pm r n \sin \psi \label{eq:10}
\end{equation}
where, for any point on the ray trajectory, $\psi$ is the angle
between the path and a radial line connecting the point and origin.
For a spherical body, the line represents a surface normal.

Below we'll calculate the fraction of emitted light escaping a
supermassive body's gravity.  Of the rays leaving a point on its
surface, only those within a narrow cone can do so. If a ray's angle
with respect to a surface normal exceeds a critical value $\psi_c$,
it returns to the body. For angles very close to this value, the
radiation goes up and orbits repeatedly, eventually returning or
escaping into space. And for smaller angles it escapes.

We can find $\psi_c$ from the photon orbit radius. For light in a
circular orbit, the wavefronts are arranged radially and their
speeds vary in proportion to their radii.  That gives the condition
\begin{equation}
\frac{\,d}{\,dr}\left( \frac{\,c}{\,r} \,\right) =\: 0
\end{equation}

Substituting for $c$ from Eq.~(\ref{eq:6}), taking the derivative,
and solving for $r$ gives the radius of the photon orbit sphere
\begin{equation}
r_{p} \:=\: 2 \mu \label{eq:12}
\end{equation}
Putting this into Eq.~(\ref{eq:8}), the refractive index $n_p$ at
the photon sphere's radius is simply
\begin{equation}
n_{p} \:=\: e
\end{equation}
The sine of the angle $\psi_c'$ for light rays in a circular orbit
is 1. Putting these values for $r_p$, $n_p$ and $\sin \psi_c'$ into
Eq.~(\ref{eq:10}) gives the conserved $k$ characterizing these light
rays, both in orbit and leaving the body's surface at the critical
angle $\psi_c$
\begin{equation}
k \:=\: 2 \mu e \label{eq:14}
\end{equation}

Rearranging Eq.~(\ref{eq:10}), $\psi_c$ is
\begin{equation}
\psi_c \:=\: \sin^{-1} \left ( \frac{\,k}{\,r_s n_s} \right )
\end{equation}
where the subscript $s$ again refers to values at the body's
surface.  Substituting for $k$ and for $n_s$ from Eq.~(\ref{eq:8})
\begin{equation}
\psi_c \:=\: \sin^{-1} \left ( \frac {\,2 \mu}{\,r_s} \: e^{1 - \,2
\mu / r_s} \right )
\end{equation}
Where $R$ is the body's radius as a fraction of the photon orbit
radius, from Eq.~(\ref{eq:12}) we can write
\begin{equation}
R \:=\: \frac{\,r_s}{\,r_p} \:=\: \frac{\,r_s}{2\mu}
\end{equation}
In terms of $R$, the previous equation becomes
\begin{equation}
\psi_c \:=\: \sin^{-1} \left ( \frac{e^{1\,-1/R}}{R} \right )
\label{eq:18}
\end{equation}

The cone of escaping light from a point on the body's surface, where
the ray angle is less than $\psi_c$, has a solid angle in steradians
given by
\begin{equation}
\Omega \:=\: 2 \pi \,( 1 - \cos \psi_c)
\end{equation}
Assuming the point radiates equally in all directions into a
hemisphere with area $2 \pi$ steradians, the fraction of emitted
light transmitted through the photon orbit is
\begin{equation}
T \:=\: \frac{\,\Omega}{\,2 \pi} \:=\: 1 - \cos \psi_c
\end{equation}
From the relation
\begin{equation}
\cos \,(\sin^{-1}x) \:=\: \sqrt{1-x^2}
\end{equation}
and Eq.~(\ref{eq:18}), the transmission as a function of $R$ is
\begin{equation}
T \:=\: 1 - \sqrt{1-\frac{e^{2- \,2/R}}{R^2}}
\end{equation}

Here are a few calculated transmissions, where the left number is
the body's radius as a fraction of the photon orbit radius:
\begin{center}
$.5 \rightarrow .32$ \\$.1 \rightarrow 7.6 \times 10^{-7}$ \\ $.05
\rightarrow 6.3 \times 10^{-15}$ \\ $.01 \rightarrow 5.1 \times
10^{-83}$
\end{center}
The radiation from a body with a radius 1 percent of the photon
orbit radius is effectively zero.  What are the implications?

Suppose a supermassive body's gravity exceeds the opposing nuclear
forces and collapse begins.  Falling inward, of course its particles
gain kinetic energy, raising its temperature and pressure. Collapse
continues as long as the body can radiate the acquired kinetic
energy.

However, the escape of radiation (including neutrinos) is cut off
before the body shrinks to 1 percent of the photon orbit radius. And
contraction stops when its gravity is counterbalanced by increased
radiation pressure. Since there is no collapse of space-time here,
the result is a darkened object of finite dimensions. Extreme
gravity is matched by extreme temperatures and pressures, with
massive particles having velocities very close to the local speed of
light.

In wave gravity, the trajectory of a massive particle with nearly
the speed of light approximates that of a light ray in a
gravitational field~\cite{kk1}. So the above equations apply
approximately to the escape of all relativistic particles from
supermassive bodies. {\em Rather than energy, the critical factor
for the escape of relativistic particles is the direction of their
trajectories.} With symmetrical gravitational fields, it's always
possible for some light or matter to escape along trajectories
aligned precisely with the field direction.

\section{Gravitational Lensing}

A radiating supermassive body appears as a dimmed disk of light (or
other radiation) larger than the body itself.  What is its size? The
disk's outer edge consists of rays leaving the body's surface at the
critical angle $\psi_c$, and we can find its apparent size from
that.  Here $\psi_c'$ will represent the angle of such rays with
respect to a line connecting an observer and the body's center. From
Eq.~(\ref{eq:10})
\begin{equation}
\sin \psi_c' \:=\: \frac{\,k}{\,r_e n_e} \:\cong\: \frac{\,k}{\,r_e}
\end{equation}
where a subscript $e$ indicates quantities at Earth's distance.

Eq.~(\ref{eq:14}) gives the conserved value of $k$ for these rays.
The apparent radius of the observed disk $r_d$ is then
\begin{equation}
r_d \:=\: r_e \sin \psi_c' \:=\: k \:=\: 2 \mu e
\end{equation}
(Again, large $e$ is the base of natural logarithms.)  Thus the
apparent size of a body inside the photon orbit sphere depends only
on its mass, not its physical dimensions.

Strictly speaking, the disk isn't an image of the body's surface.
But for each ring-shaped region of the disk, its light originates
from a corresponding circular ring on the body's surface. We can
describe a ring's position on the surface by its symmetrical angular
displacement $\theta$ from the line between the body's center and
the observer.

For light in the center of the disk, $\theta$ is zero. For the rings
of light progressively farther out, $\theta$ initially increases
gradually, then more rapidly toward the disk's edge, where it goes
to infinity. (There the rays make multiple orbits.) Thus light in
the disk's center comes from the side of the body facing the
observer, while that at the edge is a mixture of light coming from
all sides.

How does the disk's brightness vary from center to edge?  We can
find that from the relative numbers of rays emitted at different
angles, and where they appear in the disk.  We'll assume again that
points on the body's surface radiate equally in all directions, and
its luminosity is uniform everywhere.

From these assumptions, where $\psi$ is the emission angle of an
arbitrary ray (less than $\psi_c$) with respect a surface normal,
the relative numbers of escaping rays as a function of $\psi$ are
the same for the whole body as for the light emitted by a single
point. And from symmetry, this also holds for the light observed in
a disk.

Since $k$ in Eq. (\ref{eq:10}) is conserved in a spherical index
gradient, any ray observed at Earth's distance obeys the relation
\begin{equation}
r_e n_e \sin \psi' \:=\:\, r_s n_s \sin \psi
\end{equation}
Suppose two observed rays leave different points on the body's
surface, one at angle $\psi$ and another at $\psi_c$. Since the
radii and refractive indices in this equation are the same for both
rays, we can write
\begin{equation}
\frac{\,\sin \psi'}{\,\sin \psi_c'} \:=\: \frac{\,\sin \psi}{\: \sin
\psi_c}
\end{equation}
The left side of this equation represents an arbitrary ray's
apparent radial position within the disk, as a fraction of the total
disk radius.

This is the same arrangement produced by a cone of undeviated rays
from a single point source, where they strike a normal plane
surface. In that case, the light intensity in the resulting disk
decreases as the cosine of the ray angle cubed~\cite{wjs}.  In terms
of their emission angles, this must also hold for the rays in a
gravitationally lensed disk, since they have the same relative
numbers and spatial distribution.

That can be expressed as
\begin{equation}
B_\psi \:=\: B_0 \cos^3 \!\psi
\end{equation}
where $B_\psi$ is the observed brightness of rays emitted at angle
$\psi$ compared to that of a central ray $B_0$.  Eq.~(\ref{eq:18})
gives the emission angle for edge rays as a function of $R$. For an
$R$ of 0.19 or less, the disk is almost uniform, with a dimming at
the outer edge of less than 1 percent.

As shown previously~\cite{kk1}, the first-order gravitational
bending of light is the same here as in general relativity.  For a
distant background object positioned exactly behind a supermassive
body, the result would be an Einstein ring coinciding with the edge
of the above disk, at radius $r_d$.  Images of other background
objects would be displaced farther out.  Background light wouldn't
be seen within the disk.

VLBI observations of this galaxy's nucleus, Sgr A$^{\!\star}$, are
approaching the resolution needed to detect its gravitational shadow
\cite{slhz}.  In terms of background light, this theory predicts the
same shadow as general relativity. However, it's possible
gravitationally redshifted radiation from the nucleus itself may be
seen in that region.  Like other galactic nuclei, Sgr A$^{\!\star}$
is dark in visible light, but the current VLBI images show a radio
source of some kind at its approximate location \cite{slhz}.

\section{Extragalactic Jets}

Jets or other outflows are ubiquitous in most types of stars and in
active galactic nuclei (AGN).  The jets from active galactic nuclei
are typically emitted in pairs, approximately perpendicular to the
galactic planes.  Most conspicuous at radio wavelengths, these are
visible across the spectrum, sometimes including $\gamma$-rays. They
consist of plasma of undetermined composition, typically with
relativistic velocities.  And they  range in length from sub-parsec
to megaparsec scales -- as large as our local group of
galaxies~\cite{stt}.

Extragalactic jets are often highly collimated, some with opening
angles of about $1^\circ$.  Where they meet the intergalactic
medium, they balloon into broad lobes. However, the longest jets
maintain surprisingly tight collimation over most of their lengths,
and relativistic velocities are maintained almost to the ends of the
lobes.

Since galactic nuclei are thought to be black holes, the sources of
matter, energy, and driving magnetic fields for these jets have been
assumed to be accretion disks.  The jets' origins are often obscured
by an encircling dust torus. However, in cases where the point of
origin is observable, the expected accretion disks haven't been
found. Keel~\cite{wk} writes:

\begin{quote}
The ``standard" model of the central powerhouse in an active nucleus
features a very massive black hole surrounded by an accretion disk,
and it is in fact that accretion disk to which we might attribute
most of the radiation we can actually see. But, to this point, the
accretion disk has proven very shy, and observational tests for
direct signatures of the disk have come up negative or ambiguous.
This has held true with features in the optical spectrum, the
overall shape of the ultraviolet spectrum, and lines from very hot
matter seen in the X-ray spectral region. . . .

Disklike structures have indeed been seen in many nearby active
galaxies, radio galaxies and Seyferts alike. These are most often
seen as dark features from dust absorption in Hubble images, and
span diameters of 50-500 light-years. In many radio galaxies, these
disks have just the orientation we would expect - closely
perpendicular to the radio jets. There is no doubt that these are
disks, and that they are related to the central activity, but they
are not ``the" accretion disks in these objects as postulated in the
standard scheme. These disks are much too large and much too cold.
\end{quote}

M87 is a giant elliptical galaxy in the nearby Virgo cluster,
emitting a bright, well-collimated (and beautiful) jet.  From
high-resolution images made with the Keck I telescope, Whysong and
Antonucci \cite{wa} have checked for the expected \linebreak thermal
emission of an accretion disk feeding this jet.  Although the jet's
base was visible, they found it completely absent.

In response to the lack of observed disks, alternative accretion
models such as advection-dominated accretion flows (ADAF) have been
proposed. In ADAF, heating and radiation are minimized by assuming a
thin, almost frictionless plasma, moving straight toward a nucleus
and its jets from all sides, with no angular momentum.  It's not
clear how such flows would be created~\cite{sj}.

To clarify the origins of extragalactic jets, there has been a
search for specific galactic features correlating with jet activity.
However, in galaxies hosting low-luminosity AGN, no visible features
of this kind have been found. As Martini~\cite{pm} points out, the
fueling of jets from such galaxies remains an unsolved problem.

\section{Escape of Jets from Galactic Nuclei}

In wave gravity, there are no event horizons and supermassive AGN
are allowed as sources of matter, energy and magnetic fields for
jets. Again, massive particles inside galactic nuclei would have
highly relativistic velocities, and rather than their energy, their
trajectories are the critical factor for escape. For galactic nuclei
with low $R$ numbers, escape is only possible along lines following
the gravitational potential gradient almost perfectly.

If there are no magnetic fields, the trajectories of emitted
particles are unstable.  Any small deviation from the gravitational
field alignment is magnified progressively until they double back.
For charged plasma particles, which tend to follow magnetic field
lines, a magnetic field aligned with the gravitational gradient
could stabilize their trajectories.  A rotating galactic nucleus
with high internal pressure and a strong magnetic field closely
aligned with its axis of rotation would have the basic conditions
for launching jets.

It's believed most or all luminous galaxies have supermassive
objects at their centers. For low and intermediate-luminosity
galaxies these tend to be compact stellar nuclei, while for massive
galaxies they're assumed to be black holes.  The term Central
Massive Object (CMO) is used for both. Ferrarese {\em et
al.}~\cite{lf} find the CMO and host galaxy masses are tightly
coupled, with the former about $0.2$ percent of the latter. They
add, ``Unfortunately, the physical mechanisms underlying this
connection remain obscure."

There is also increasing evidence that galaxies are cyclical and
that most may have active phases.  Martini~\cite{pm} has proposed
luminous disk galaxies might appear successively as Seyfert, LINER
(Low-Ionization Nuclear Emission Region) and inactive galaxies,
where the relative numbers of these types suggest duty cycles of 10,
30, and 60 percent respectively. The total cycle time would depend
on the mass of the nucleus, estimated at $10^8$ years for one of
about $10^7$ solar masses.

Wave gravity allows a new galaxy model consistent with this picture.
Some of its features are inherited from an early model of radio
galaxies proposed by Alfv\'{e}n~\cite{ha}.  The large jets from
radio galaxies resemble fountains. Where a collimated jet contacts
the intergalactic medium, it spreads outward and falls back toward
the central galaxy, in a ``cocoon" surrounding the outgoing jet.
Much of this matter is recaptured by the galaxy and may eventually
accrete to the nucleus again in some form.

In Alfv\'{e}n's model, this circulation results from electromagnetic
forces powered by a galaxy's rotation.  A magnetic field
perpendicular to its rotational plane would polarize orbiting
plasma, giving a voltage difference between the galaxy's center and
periphery.  Where $\bf B$ is the magnetic field and $v$ the plasma's
orbital velocity, a plasma particle with charge $q$ experiences a
magnetic Lorentz force in the radial direction
\begin{equation}
{\bf F}\:=\: q \left(\frac{{\bf v} \times {\bf B}}{c} \, \right)
\end{equation}
In the case where B is opposite the galaxy's spin vector, plasma
ions are driven preferentially toward the center of the galaxy and
electrons outward. Such a system is called a ``unipolar inductor,"
or ``homopolar generator."

The resulting voltage drives plasma circulation, which carries
electric current outward from the central galaxy in the axial jets,
diverges, and returns to the outer galaxy to complete the circuit.
Alfv\'{e}n predicted pinched currents, carried by one or more
twisted filaments. Such filaments are observed now in
high-resolution images of jets in nearby radio
galaxies~\cite{ohc,lz}, and in their lobes~\cite{pdc,ebf,dac}.

According to Alfv\'{e}n~\cite{ha}, solar system measurements imply
the rotating Sun acts as a unipolar inductor.  A popular model of
pulsar jets, due to Goldreich and Julian~\cite{gj}, also treats
those bodies as unipolar inductors. To explain AGN jets in the
context of general relativity, Blandford and Znajek~\cite{bz} have
proposed a model where plasma inside the ergosphere of a rotating
black hole acts as a unipolar inductor.  Thorne, Price and
MacDonald~\cite{tpm} find the event horizon of a black hole would be
electrically conductive and could serve the same function,
propelling extragalactic jets fueled by accreting matter.

A rotating AGN could behave as a unipolar inductor for jets in wave
gravity.  In this case, the reservoir of matter available for
fueling is much larger than an accretion disk: it's the AGN itself.
Given the possibility of very massive jets, gravity might assist
their collimation, and the formation of observed knots. In addition,
gravitational contraction becomes a possible energy source for X-ray
hot spots coinciding with the knots.  Kataoka and Stawarz~\cite{ks}
note some of these are more intense than allowed by current models.

Prominent hot spots are also seen near the ends of radio galaxy
jets. Alfv\'{e}n attributes these to plasma double layers, forming
where the plasma flow diverges into the lobes. Radio polarization
measurements of a jet in 3C219 by Clarke {\em et al.}~\cite{dac}
show its magnetic field lines match the flow there. And recent
laboratory experiments by Sun {\em et al.}~\cite{xs} confirm double
layers form in plasmas at the point a diverging magnetic field is
encountered.

As Alfv\'{e}n describes, the effect of such double layers would be
to accelerate high-velocity electrons inward toward the central
galaxy, and high-velocity ions outward as cosmic rays.  While the
latter travel great distances, migration of the electrons would be
limited, mainly by their energy loss to synchrotron radiation. Here
we note this would remove ions from the general galaxy region,
leaving plasma there with a net negative charge.

What does this do to a galactic nucleus?  In accord with the
emerging view, we'll suppose its jets are intermittent on long time
scales. During a quiescent phase, as negatively charged plasma
accretes to the nucleus, negative charge is concentrated there.  A
supermassive nucleus could capture electrons as long as its gravity
exceeds the electrical repulsion. And its negative charge would
strengthen continuously as long as the electron capture rate exceeds
that for positive charges.

Unlike a black hole, here unbalanced charge in a spinning
supermassive nucleus would create a magnetic field extending outside
it.  (Without curved space-time, this is no different than in
ordinary electrodynamics.) We'll assume a nucleus is surrounded by
plasma orbiting in roughly the same direction it spins. Where the
net charge is negative and $\bf B$ is the resulting external
magnetic field, the radial ${\bf v} \times {\bf B}$ force on plasma
particles in the equatorial region would be directed toward the
nucleus for electrons, and away for ions.

Preferential accretion of electrons is also favored by their orbital
decay from synchrotron radiation, which is faster than for ions.
Given galactic nuclei can concentrate electric charge, and the
likelihood of extreme rotational velocities, the possible magnetic
fields appear much larger than those obtainable from existing
models.

Jets would be launched when the magnetic field becomes strong enough
to guide plasma escape.  To account for the self-collimation of
astrophysical jets, Honda and Honda~\cite{hh} have found a need for
excess plasma electrons. Those are inherent here. The resulting
current gives a toroidal magnetic field helping confine a jet.
(While plasma flows in the same direction, the current direction is
opposite that in Alfv\'{e}n's radio galaxy model.)

Since excess charge on a body is concentrated near its surface,
these jets would be expected to carry away electrons in
disproportionate numbers. That would allow the negative charge on a
galactic nucleus to dissipate, leading eventually to loss of its
magnetic field and cessation of the jets. At that point, the nucleus
starts accumulating matter and charge again, and the cycle repeats.

Burbidge~\cite{gb} argues present models don't account properly for
the huge radiated energies of large extragalactic jets. Some exceed
the total radiated outputs of their source galaxies by over two
orders of magnitude.  In this model, since jets originate from
supermassive AGN instead of accretion disks, their possible energies
are larger also.

\section{Matter Recycling}

Even in the standard model of jets from accretion disks, there is
some gravitational recycling of matter.  Kinetic energy gained
allows the spent nuclear fuel of accreting neutron stars to
dissociate into constituent particles.  And while most of the
resulting plasma is lost to black holes, some fraction of it fuels
jets, giving rise to hydrogen for new stars.

More substantial recycling is allowed here. For supermassive
galactic nuclei with low $R$ numbers, which radiate very little,
most of the kinetic energy acquired by infalling matter is retained.
Since that energy remains available for pressurizing jets, much of
the matter entering a nucleus may escape eventually as plasma.

Fragile {\em et al.}~\cite{pcf} write:  ``Accumulating observational
evidence for a number of radio galaxies suggests an association
between their jets and regions of active star formation."  They
describe various examples of star-forming regions near radio galaxy
lobes. Their finding is that those can be attributed to the
destabilizing effects of jets on pre-existing clouds in the
intergalactic medium. From the more massive jets allowed in this
model, it's also conceivable their filaments provide the
star-forming matter.

Seyferts are active galaxies, usually disks, with bright central
regions.  The central brightness is due to AGN jets, circumnuclear
rings of young stars, or often both~\cite{rcf}. The origin of this
``starburst-AGN connection" isn't well understood.  Based on the
standard model of AGN as black holes, it's suggested that outflows
from the Seyfert starbursts accrete and fuel the jets. But it isn't
clear how the former are fueled. It's also proposed that both may
result from galaxy mergers. However, a survey by Knapen~\cite{jhk}
finds: ``There is no convincing evidence that AGN hosts are
interacting more often than non-AGN."

Compared to radio galaxy jets, those from Seyfert galaxies are
generally much shorter and less collimated, and could supply matter
directly to the central galaxy region.  Here it's possible the
starbursts are fueled by massive jets from their nuclei.

A small-scale starburst also surrounds Sgr A$^{\!\star}$ at the
center of this galaxy.  The innermost stars and their orbits have
been surveyed to find the central body's mass (about $4 \times 10^6$
solar masses) and precise location.   Absorption spectra for one of
these, S0-2, were first obtained by Ghez {\em et al.}, who
identified it as a young main-sequence star. This result is a
``paradox of youth." They write:

\begin{quote}
It is challenging to explain the presence of such a young star in
close proximity to a supermassive black hole. Assuming that the
black hole has not significantly affected S0-2's appearance or
evolution, S0-2 must be younger than 10 Myr and thus formed
relatively recently. If it has not experienced significant orbital
evolution, its apoapse distance of 1900 AU implies that star
formation is possible in spite of the tremendous tidal forces
presented by the black hole, which is highly unlikely. If the star
formed at larger distances from the black hole and migrated inward,
then the migration would have to be through a very efficient
process. Current understanding of the distribution of stars,
however, does not permit such efficient migration.
\end{quote}

Of the approximately 80 massive stars concentrated in the central
parsec of this galaxy, a recent study by Paumard {\em et
al.}~\cite{tp} finds almost all belong to two coherent systems,
orbiting on separate planes.  Neither plane corresponds to the
Galaxy's, and the two are roughly orthogonal.  Seen from the North
Galactic Pole, stars in the innermost system orbit clockwise, while
those in the ring-like system encircling that go counter-clockwise.
Their study also finds: ``The stellar contents of both systems are
remarkably similar, indicating a common age of $\simeq 6 \pm 2$
Myr."

As Paumard {\em et al.} argue, even in the presence of strong tidal
forces near  Sgr A$^{\!\star}$, it's plausible these stars were
formed there, if interstellar matter in the region were much denser
in the past. Beyond the existence of young stars very near the
Galaxy Center, a dual-jet origin might explain their uniform ages
and why they're found in two systems with skewed orbital planes.

\section{Conclusions}

On the assumption general relativity is correct, galactic nuclei are
often called ``black holes."  However Einstein~\cite{ae} insisted
there are none: ``the Schwarzschild singularities do not exist in
physical reality."  And for no astronomical body has clear evidence
of an event horizon been found.

If these objects are black holes, experiencing only one-way
accretion of matter, how do nuclei in old galaxies maintain the same
fraction~\cite{lf} of galactic mass as in young ones? In this
gravity theory, escape of accumulating matter is possible via
periodic jets. Some gravitationally redshifted light can also
escape, and may be observable within the shadow region predicted by
general relativity for Sgr A$^{\!\star}$.

Preceding papers~\cite{kk1,kk2} described a number of problems with
general relativity which don't arise in wave gravity. We can add
there is no ``problem of information loss" or ``mass maintenance
problem" for galactic nuclei, and no ``fueling problem" for their
jets.

\section{Acknowledgements}

I thank Stan Robertson and Howard Drake for helpful discussions.

\vfill\eject

\end{document}